\documentclass[pdftex,12pt]{article}
\usepackage{jheppub,amsmath,amsfonts,amsbsy,latexsym,jheppub,amssymb,amscd,amstext}
\usepackage{pst-node}
\usepackage{dsfont}
\usepackage{overpic,youngtab}
\pdfoutput=1
\usepackage{rotating}
\usepackage{tikz}
\usepackage{slashed}

\def\be{\begin{equation}}
\def\ee{\end{equation}}
\def\bea{\begin{eqnarray}}
\def\eea{\end{eqnarray}}
\def\pd{\partial}
\def\a{\alpha}
\def\b{\beta}

\def\g{\gamma}
\def\d{\delta}
\def\m{\mu}
\def\mp{\m^\prime}
\def\np{\n^\prime}
\def\n{\nu}

\def\l{\lambda}
\def\lp{\lambda^\prime}

\def\r{\rho}

\def\cD{\cal D}

\def\xp{x^\prime}

\def\s{\sigma}

\def\bma{\begin{pmatrix}}
\def\ema{\end{pmatrix}}

\def\bi{\begin{itemize}}
\def\ei{\end{itemize}}

\begin{document}

		\vspace*{-1cm}
		\phantom{hep-ph/***} 
		{\flushleft
			{{FTUAM-20-2}}
			\hfill{{ IFT-UAM/CSIC-20-12}}}
		\vskip 1.5cm
		\begin{center}
		{\LARGE\bfseries CFT in Conformally Flat Spacetimes}\\[3mm]
			\vskip .3cm
		
		\end{center}
		\vskip 0.5  cm
		\begin{center}
			{\large Enrique Alvarez and Raquel Santos-Garcia }
			\\
			\vskip .7cm
			{
				Departamento de F\'isica Te\'orica and Instituto de F\'{\i}sica Te\'orica, 
				IFT-UAM/CSIC,\\
				Universidad Aut\'onoma de Madrid, Cantoblanco, 28049, Madrid, Spain\\
				\vskip .1cm

				\vskip .5cm
				\begin{minipage}[l]{.9\textwidth}
					\begin{center} 
						\textit{E-mail:} 
						{enrique.alvarez@uam.es}, {raquel.santosg@uam.es}
					
					\end{center}
				\end{minipage}
			}
		\end{center}
	\thispagestyle{empty}
	
\begin{abstract}
	\noindent
	A new class of conformal field theories is presented, where the background gravitational field is conformally flat.
	Conformally flat (CF) spacetimes enjoy conformal properties quite similar to the ones of flat spacetime. The conformal isometry group is of maximal rank and the conformal Killing vectors in conformally flat coordinates are {\em exactly} the same as the ones of flat spacetime. In this work, a new concept of distance is introduced, the {\em conformal distance}, which transforms covariantly under all conformal isometries of the CF space. It is shown that precisely for CF spacetimes, an adequate power of the said conformal distance is a solution of the non-minimal d'Alembert equation.
\end{abstract}

\newpage
\tableofcontents
	\thispagestyle{empty}
\flushbottom

\newpage
	\setcounter{page}{1}
\section{Introduction}
The importance of flat spacetime Conformal Field Theories (CFT) in theoretical physics can be hardly exaggerated (cf. for example, \cite{Rychkov,Osborn}). They are fixed points of the renormalization group, and as such, they are in some sense, the simplest of all quantum field theories. It is clear than the more examples we have to analyze their  physical behavior, the better.
\par
The present paper aims to generalize this whole setup of Conformal Field Theories to a particular instance of curved spacetimes, namely to conformally flat (CF) spacetimes. 
CF spacetimes can be characterized in an invariant way as those which are Weyl-flat (the Weyl tensor vanishes). They correspond to Type 0 in Petrov's classification of Einstein spacetimes \cite{Petrov}; we shall correspondingly denote them as {\em Weyl-flat} spacetimes. The Weyl tensor is the traceless piece of the Riemann tensor
\bea
&W_{\a\b\g\d}\equiv R_{\a\b\g\d}-{1\over n-2}\left\{g_{\a\g} R_{\b\d}-g_{\a\d} R_{\b\g}-g_{\b\g} R_{\a\d}+g_{\b\d} R_{\g\a}\right\}+\nonumber\\
&+{1\over (n-1)(n-2)}\,R\,\left\{g_{\a\g} g_{\d\b}-g_{\a\d} g_{\g\b}\right\}
\eea
When the Weyl tensor vanishes, Riemann's tensor can be fully expressed in terms of the metric, the Ricci tensor, and the scalar curvature.
It can be then proved \cite{Eisenhart} that  there is a local system of coordinates such that the metric is conformally flat, that is,
\be
ds^2=a^2(x) \eta_{\m\n} dx^\m dx^\n.
\ee
where $\eta_{\m\n}$ is the Minkowski metric.
Quantum field theories in curved spacetimes are generically non-renormalizable because new dimension four operators involving the gravitational field are generated by quantum effects. It would be interesting to check whether the implementation of conformal symmetry generates consistent QFT.
\par
The set of conformally flat spacetimes is quite large. It includes in particular, all Friedmann-Robertson-Walker (FRW) spacetimes, of interest in Cosmology. Both de Sitter and anti-de Sitter spacetimes are particular cases in which the curvature is constant. In \cite{Backovsky} there is a quite complete classification of these spacetimes.
\par
The plan of the paper is as follows. Section 2 is devoted to the exploration of the group of conformal isometries of CF spacetimes (which is of maximal dimension), where we find a simple relation between the conformal Killing vectors of flat spacetimes and the ones of CF spacetimes. In section 3, we discuss a new concept of distance, which although related to the geodesic distance (Synge's world function) is not identical to it. This distance, dubbed by us as {\em conformal distance} transforms covariantly under all conformal isometries of the CF space, and gets a very simple expression in the natural coordinate system. In the next sections, we explore the possibility that some power of the said conformal distance is a solution of the d'Alembert equation. The answer is that only when the Weyl invariant non-minimal coupling is introduced, is this possible at all, and in this case, the explicit expressions are similar to the ones in flat spacetime. Dirac fermions are explored next, and the simplest violations of conformal invariance owing to interactions are also deduced in a simple manner. We end up with our conclusions and suggestions for future work.

\section{The group of conformal isometries of CF spacetimes.}
We want to show that there is a simple relationship between the well-known conformal Killing vectors (CKV)  of flat spacetime
\be
\pd_\m k_\n+\pd_\n k_\m={2\over n} \pd_\l k^\l\,\eta_{\m\n}
\ee
(with $
k^\m\equiv \a^\m+\l x^\m +\omega^\m\,_\n x^\n+2(x.\b) x^\m-\b^\m x^2$ and $x^2\equiv \eta_{\m\n} x^\m x^\n$), and the ones of a conformally flat (CF) spacetime with metric $g_{\m\n}= a^2 (x)\eta_{\m\n}$. The definiton for a CKV of a generic curved spacetime with metric $g_{\m\n}$ reads
\be
\nabla_\m \xi_\n+\nabla_\n\xi_\m={2\over n} \nabla_\l \xi^\l\,g_{\m\n}\equiv w(x)g_{\m\n},
\label{KVcurved}
\ee
where the definition of $\omega(x)$ as the piece multiplying the metric in the rhs of the equation will be relevant in subsequent definitions. Although finding the CKV of a generic metric is not an easy task, one can see that in the case of CKV for CF spacetimes they are simply related to the ones of flat spacetime.
\par
For CF spacetimes the connection has the simple form
\be
\Gamma^\m_{\n\r}=-{\eta^{\m\l}\pd_\l a\over a} \eta_{\n\r}+{\pd_\n a\over a} \d^\m_\r+{\pd_\r a\over a}\d^\m_\n,
\ee
and we can then write
\be
\nabla_\m\xi^\m=\pd_\m\xi^\m+{n\over 2}\xi^\l\pd_\l\log\,a^2.
\ee
The equation we have to solve is then\footnote{Let us note that we have to be careful with the factors of $a$ appearing when rising and lowering indices. For example in \eqref{2.5}  we have $\xi^\mu = a^{-2} \eta^{\m \n} \xi_\n $. We explicitly write the Minkowski metric to avoid confusion when necessary.}
\be
\pd_\m \xi_\n+\pd_\n  \xi_\m-2 \xi_\l\left(-{\eta^{\l\s}\pd_\s a\over a}\eta_{\m\n}+{\pd_\m a\over a} \d^\l_\n+{\pd_\n a \over a}\d^\l_\m\right)={2\over n}\left(\pd_\l \xi^\l+n {\xi^\s\pd_\s a\over a}\right)a^2 \eta_{\m\n}
\label{2.5}
\ee
We claim that the ansatz
\be
\xi_\m\equiv  a^2 k_\m  \Longleftrightarrow \xi^\m=k^\m
\ee
does the job. This means that the CKV are exactly the same, generating a conformal algebra of dimension $D=\dfrac{(n+1)(n+2)}{2}$, namely,  $SO(2,n)$.
\par
To summarize, the isometry group of n-dimensional Minkowski spacetime (generated by all Killings) is $ISO(1,n-1)\sim SO(1,n)$, which is (a contraction of the) de Sitter group. Anti de Sitter $SO(2,n-1)$ has the same number of generators; they are different real forms of $SO(n+1)$. All those spacetimes have isomorphic conformal group, generated by the CKV, namely, some real form of $SO(n+2)$, like $SO(2,n)$. In fact, starting from the flat space equation
\be
\left\{\pounds(k)\eta_{\m\n}={2\over n}\pd_\l k^\l  \eta_{\m\n}\right\}\equiv\left\{\pd_\m k^\l \eta_{\l\n}+\pd_\n k^\l \eta_{\l\m}={2\over n}\pd_\l k^\l \eta_{\m\n}\right\},
\ee
we easily deduce
\be
\left\{\pounds(\xi)g_{\m\n}={2\over n}\nabla_\l \xi^\l\ g_{\m\n}\right\}\equiv \left\{\pd_\m\xi^\l g_{\l\n}+\pd_\n\xi^\l g_{\l\m}={2\over n}\pd_\l\xi^\l g_{\m\n}\right\}.
\ee
\par
It is worth noticing that  $\pd_\l k^\l=0$ (flat spacetime Killings) does not imply $\nabla_\l \xi^\l=0$ 
so that the $\frac{n(n+1)}{ 2}$ flat space Killings generating the  $ISO(1,n-1)$ algebra are {\em not} assymptotically flat Killings in general. Conversely, the ${n(n+1)\over 2}$ Killings generating the isometry algebra $SO(1,n)$ of the CF spacetimes (that is, $\nabla_\l\xi^\l=0$) are just  CKV of flat spacetime in general (because then $\pd_\l k^\l\neq 0$).
%
%
 \section{Conformal distance}
Now that we know the simple relation between the CKV of flat spacetime and the ones of CF spacetimes, let us further explore other properties of conformal field theories in this kind of spacetime.
\par 
The general definition of Synge's  world function \cite{Synge} is 
\be
\s(x,\xp)\equiv {\lp-\l\over 2}\int_x^{\xp} g_{\a\b}[x(\l)] \,{dx^\a\over d\l}\,{d x^\b\over d\l} d\l={1\over 2} s_{x,\xp}^2
\ee
where the integral is done along a geodesic $x^\m(\l)$ (assumed to be unique and timelike) that joins both points. It is essentially the square of the geodesic arc length, and is so defined as to be positive semi-definite in pseudo-Riemannian spacetimes.
This definition is the only one available in the generic manifold without any Conformal Killing Vectors (CKV).
The world function is not easy to compute explicitly, except in some simple cases, such as constant curvature spacetimes, which are included in Appendix \ref{A}. In general spacetimes, all that can be done is to compute coincidence limits of its derivatives \cite{DeWitt}.
\par
When the spacetime is such that there exists a nonvanishing set of CKV, there is another definition however, which coincides with a function of the former in flat spacetime as well as in spacetimes of constant curvature (cf. Appendix \ref{A}). This new definition determines the {\em conformal distance} from the physically desirable conditions
\be
\left(\pounds_{k(x)}+\pounds_{k(\xp)}\right)\, C(x,\xp)={w(x)+w(\xp)\over 2}\, C(x,\xp)\equiv \omega(x,\xp) C(x,\xp)
\label{conformaldistance}
\ee
which must hold for all CKV, $k_\m$. This condition demands that the action of the CKV on the conformal distance is the natural one. The function $w(x)\equiv {2 \over n} \nabla_\l \xi^\l$ appearing in this definition is defined in \eqref{KVcurved}.
It could have been expected that the geometric concept of world function also satisfies these criteria in our case so that our two definitions of distance coincide. The surprising thing is that it does not; at least not exactly.
\par
Let us compute now the conformal distance in our case. The defining equation \eqref{conformaldistance} leads to
\be
\left\{k^\m(x)\pd_\m+k^\m(\xp)\pd_{\mp}\right\}\log{C(x,\xp)\over a(x)a(\xp)}=2\l-2 \eta_{\m\n} \b^\m \left(x^\n-x^{\np}\right)
\ee
which in turn fully determines the conformal world function to be
\be
C(x,\xp)=a(x) a(\xp)(x-\xp)^2
\label{C}
\ee
where $x^2\equiv \eta_{\m\n} x^\m x^\n$.

We have then been able to determine the precise form of the conformal distance for all CF spacetimes. In Appendix \ref{A} it has been worked out in detail the expression for the world function in the particular case of constant curvature spacetimes \eqref{geodesicdistance} and shown not to be equal to the conformal distance \eqref{conformaldS}. The two concepts, although related, are then not exactly equivalent.
\section{Conformal fields in CF spacetimes}
The paradigm of a conformal field is the free massless scalar field in four-dimensional Minkowski spacetime, where correlators are determined essentially by dimensional analysis,
\be
\langle 0_-\left|\phi(x)\phi(\xp)\right|\rangle={1\over \left(x-\xp\right)^2}.
\ee
Let us try to generalize this idea to our case. After computing the form of the conformal distance for CF spacetimes, we can see whether the solution of the d'Alembertian equation can be written as some power of this particular distance. 
\par
The d'Alembertian for conformally flat spacetimes reads
\be
\Box_{\text{\tiny{CF}}}\equiv {1\over a^n}\pd_\m\left(a^{n-2}\eta^{\m\n} \pd_\n\right)=(n-2){\pd_\m a\over a^3}\eta^{\m\n}\pd_\n+ {1\over a^2}\eta^{\m\n}\pd_\m\pd_\n
\label{boxCF}
\ee
Before analyzing its action on functions of the conformal distance, let us compute the Minkowskian d'Alembertian acting on the conformal distance \eqref{C}. After some computations one can see that 
\bea
\Box C(x,\xp)\equiv\eta^{\m\n} \pd_\m\pd_\n C(x,\xp)&=& \Box a(x) a(\xp)(x-\xp)^2+2 n  a(x) a(x^\prime) \nonumber \\ 
&&+4  (x-\xp)^\m\pd_\m a(x) a(x^\prime).
\label{BoxC}
\eea
But we are interested in its action on functions that only depend on the conformal distance, $G\left[C(x,\xp)\right]$, so that 
\be
\Box G=G^{\prime\prime}(\pd_\m C)^2+ G^\prime \Box C
\ee
where $G' \equiv  \frac{\pd G\left[C(x,\xp)\right]}{\pd C(x,\xp)}$. 
\par
We can then write the action of the CF d'Alembertian on functions of the conformal distance in terms of the action of the one in Minkowski spacetime
\be
\Box_{\text{\tiny{CF}}} G\left[C(x,\xp)\right]={1\over a^2}\left[G^{\prime\prime}(\pd_\m C)^2+ G^\prime \Box C\right]+(n-2)G^\prime {\pd^\m a\over a^3}\pd_\m C
\ee
Using the form of \eqref{BoxC} and denoting $a^\prime\equiv a(\xp) $ in order to shorten the writing, we finally get
\bea
\Box_{\text{\tiny{CF}}} G\left[C(x,\xp)\right]&=&
{G^\prime\over a^2} \left[\left({\Box a\over  a} C + 2 n a a^\prime+ 4 (x-\xp)^\m \pd_\m a a^\prime\right) \right. \left. +(n-2){\eta^{\m\l} \pd_\l a\over a}\left({\pd_\m a\over a }C \right. \right.\nonumber\\
&& \left. \left.+ 2 a a^\prime (x-\xp)_\m\right)\right] +{1\over a^2} G^{\prime\prime}\left({\pd_\m a\over a} C+ 2 a(x) a(\xp) (x-\xp)_\m\right)^2. \nonumber \\
\eea
In order to check whether there is a solution that depends on the conformal distance only, the terms involving $(x-\xp)^\m \pd_\m a$ ought to  cancel by themselves. This imposes the condition
\be
4{a(x) a(\xp)\over a^3} G^{\prime\prime} C+{4 a(\xp)\over a^2} G^\prime+(n-2) 2 { a(\xp)\over a^2} G^\prime=0.
\ee
That is
\be
2 G^{\prime\prime} C(x,\xp)+ n G^\prime=0,
\label{eqG}
\ee
which completely determines the function $G[C]$ up to two arbitrary constants  $A_1$ and $A_2$
\be
G\left[C(x,\xp)\right]=A_1\, \left(C(x,\xp)\right)^{1-{n\over 2}}+A_2.
\ee
The remaining terms then read
\bea
\Box_{\text{\tiny{CF}}} G&=&{1\over a^2} G^\prime\left({\Box a\over a}+{n-4\over 2 a^2} (\pd_\m a)^2\right) C(x,\xp) = -\dfrac{R}{2(n-1)} \, G' C(x,\xp)=\nonumber\\
&=&{n-2\over 4 (n-1)}\,R\,\left(G-A_2\right),
\eea
and we can always redefine $G\rightarrow G-A_2$. Finally we find that 
\be
\left[\Box_{\text{\tiny{CF}}}-{n-2 \over 4 (n-1)}\, R \right]\,G\left[C(x,\xp) \right]= 0.
\ee
This equation conveys the fact that in order to get Green functions that depend on the conformal distance only we need to consider the Weyl covariant operator. Let us insist in the remarkable fact that when expressed in terms of the conformal distance, the Green's functions for conformal fields are exactly the same as the ones in flat spacetime. The only fact is that these conformal fields are necessarily non-minimally coupled to the gravitational field.

\section{Dirac fermions}
In this section, we carry on the same analysis for fermions (particularized for dimension 4), which necessarily are defined in an inertial frame. In conformally flat coordinates the simplest coframe is given by $e^a_\m= a(x) \d^a_\m$. We employ  the standard notation $\pd_a f(x)\equiv \vec{e_a}\,(f)\equiv e_a^\l\pd_\l f(x)$
 \footnote{The structure constants are defined through $\left[\vec{e}_a,\vec{e}_b\right]=C_{ab}^c \vec{e}_c$
	and are given by $C_{ab}^c= {\d_a^c\,  \pd_b a -\d_b^c\, \pd_a a \over a^2}$.
	It follows that the  spin connection reads
	\be
	\omega_{ a|bc}\equiv {1\over 2}\left(C_{ac|b}+C_{ba|c}+C_{bc|a}\right)={\pd_c a(x)\over a^2}\,\eta_{ a b}-{\pd_b a(x)\over a^2}\,\eta_{ ac}.
	\ee
	In order to project to spacetime indices, one has to be careful with the fact that
	\be
	\pd_a a(x)\equiv {\pd a(x)\over \pd x^a}\neq e_a^\m \pd_\m a(x)=a\pd_a a(x)
	\ee
}.
The spin connection in CF spacetimes takes the form
\be
\omega_{\m|bc}\equiv e_\m^a \omega_{a|bc}={\pd_c a(x)\over a}\,\eta_{ \m b}-{\pd_b a(x)\over a}\,\eta_{ \m c}
\ee.
With this we can construct the action principle for a  Dirac fermion in a Weyl-flat background, which is given by
\be
S\equiv\int d^4 x\, a^4\,\bigg\{i \overline{\psi} \, \g^a e_a^\m {\cal D}_\m \psi\bigg\},
\ee
where the conformal factor $a^4$ comes from the volume factor. The derivative action on the fermion field is defined as
\be
{\cal D}_\m\equiv \nabla_\m + \Gamma_\m= \pd_\m+\omega_\m^{ab} \Sigma_{ab}
\ee
with $\Sigma_{ab}\equiv {1\over 4} \g_{ab}={1\over 8}\left[\g_a,\g_b\right]$. 
The piece involving the spin connection reads
\be
{1\over 4}\g^a e_a^\m\omega_{\m| b c}\g^{bc}\equiv {1\over 4}\g^a\omega_{a| b c}\g^{bc}={1\over 4 a^2}\g^a\g^{bc}\left(\eta_{ab}\, \pd_c a-\eta_{a c}\, \pd_b a\right)= {3\over 2} {\g^c \pd_c a\over a^2}
\ee
where we have used the fact that $\g^a \g_{ab}= 3 \g_b$.
\par
As in the previous case, we want to explore the Green's function that behaves in a easy way in CF spacetimes. We can compute the Dirac propagator which obeys
\be
\slashed{\cD}\,S(x,\xp)\equiv i  \, \g^a e_a^\m  {\cal D}_\m S(x,\xp)\equiv i \g^\m\left({1\over a} \pd_\m+{3\over 2 a^2}\pd_\m a\right) S(x,\xp)=0,
\ee
where all indices are contracted with the flat metric (all the conformal factors have already been fully revealed) and the gamma-matrices from now on are also the ones corresponding to flat spacetime.
This equation can be written as
\be
i\slashed{\pd}\left(a(x)^{3\over 2} S(x,\xp)\right)=0.\label{diracprop}
\ee
This means that
\be
\Box\left(a(x)^{3\over 2} S(x,\xp)\right)=0.\label{}
\ee
Recall that $\Box$ is just the flat space wave operator. Equation \eqref{diracprop} is equally fulfilled when differentiating at point $x'$. In order to incorporate this fact,  we can multiply \eqref{diracprop} by the corresponding factor of $a(x')$ as it is invisible to derivatives at the point $x$. We can then write the solution to \eqref{diracprop} at both points as
\be
a(x)^{3 \over 2} a(x')^{3 \over 2} S(x,\xp)=i\slashed{\pd}{1\over (x-\xp)^2}=-2i  {\slashed{x}-\slashed{\xp}\over (x-\xp)^4}.
\ee
An appealing solution is given by
\be
S(x,x')=\g^a S_a(x,x'),
\ee
where\footnote{The factors and the sign are just a matter of convention in the definition \cite{Freedman}.}
\be
S_a(x,x') = -\dfrac{1}{2\pi^2} {a(x)^{1/2} a(\xp)^{1/2}\over C(x,\xp)^2}(x-\xp)_a=-\dfrac{1}{2\pi^2}{ (x-\xp)_a\over a(x)^{ 3/2} a(\xp)^{ 3/2}(x-\xp)^4}.
\ee
%
%
%
%
%
We conclude that elementary fermionic solutions can again be easily expressed in terms of our conformal distance. 
\par 
	To conclude, let us comment on the square of Dirac's operator in an arbitrary gravitational background. The well-known \cite{Lichnerowicz,Lichne} Weitzenb\"ock-Lichnerowicz identity on the square of this operator (which is the only one that is Weyl conformally invariant) reads
	\bea
	&\g^\a{\cal D}_\a\g^\b{\cal D}_\b=\g^\a\g^\b{\cal D}_\a{\cal D}_\b=g^{\a\b} \cD_\a \cD_\b+\left[\g^\a,\g^\b\right]\cD_\a\cD_\b=\Box^{\text{s}}+\nonumber\\
	&+\left[\g^\a,\g^\b\right]\left(\nabla_\a \Gamma_\b-\nabla_\b \Gamma_\a+\left[\Gamma_\a,\Gamma_\b\right]\right)=\Box^\text{s}-{1\over 4}\left[\g^\a,\g^\b\right] R_{\a\b ab}\g^{ab}.
	\eea
	The algebraic Bianchi identity now implies that $\g^\m\g^\n\g^\r\g^\s\,R_{\m\n\r\s}=2 R$ and it follows that we can write
 \be\left(\g^\a{\cal D}_\a-m\right)\left(\g^\b{\cal D}_\b+m\right)=\Box^\text{s}-m^2-{1\over 4}R. \ee
	It is interesting to see that the mass term in $(\g^\a{\cal D}_\a-m)$ can be so arranged that the Dirac spinor obeys the Weyl invariant formula (in $n=4$) $
	\left(\Box-{1\over 6}R\right)\psi=0$ \cite{Shore}.

\section{CFT correlators}
Conformal symmetry is generically broken by quantum effects once interactions amongst the fields are taken into account. This is because the regularization process introduces a mass scale, and some reminders of it are almost always present in the renormalized theory as well.
Nevertheless, there are theories (usually with at least 8 supersymmetric conserved charges) in which those anomalies are not present.
\par
Our aim in this section is to show how the breaking of Weyl invariance due to quantum interactions can be computed in CF spaces as easily as it is computed in flat spacetime.
\par
It is, of course, possible to renormalize any QFT in curved space using dimensional regularization in momentum space (cf. for example \cite{Jack}).
In CFT in general and CF spacetimes in particular, however, it is advantageous to work in position space, owing to the simplicity of the propagators when so expressed.
Actual calculations are very similar to flat spacetime ones. Let us examine a couple of examples to illustrate this point. Consider the conformal $\l \phi^4$ theory in $n=4$ dimensions in a Weyl-flat space. We have seen that the propagator in CF spacetime will have the form
\be\Delta(x,\xp)= \dfrac{1}{4\pi^2}{1\over a(x)a(\xp) (x-\xp)^2}.
\ee
This two-point function is exact in the free case; that is, it includes all corrections coming from the existence of a gravitational background.
\par
When treating the theory in position space, all the tricks of differential renormalization (DR)\cite{Freedman} \cite{Gracia} can then be used with only slight modifications. Namely, the flat spacetime distance $ (x-x')^{2}$ is replaced by $a(x)a(x')(x-x')^{2}$. The key point is that the singular 
coincidence limit $x\rightarrow x'$  of the conformal distance can be trivially computed from the corresponding limit in the Minkowskian distance. All diagrams in CFT in CF spacetimes can then be obtained
from the flat spacetime ones just by inserting factors of $a(x)$ and $a(x')$ in the adequate place.
\par
To be specific, let us examine corrections to the two-point function coming from the self-interaction. The only diagram to one loop is the tadpole, which vanishes in DR. The two-loop contribution is given in \cite{Freedman} by
\be
\Delta^{(2)}(x,\xp)=-{\l^2\over 6}  \left(\Delta\left(x,\xp\right)\right)^3. \label{propagator}
\ee
Introducing the propagator \eqref{propagator} for CF spacetimes we get
\be
\Delta^{(2)}(x,\xp)=-{\l^2\over 384 \pi^6} \, \dfrac{1}{a(x)^3 a(x')^3(x-x')^6},\label{twopoint}
\ee
where the only difference with the flat spacetime case is encoded in the conformal factors $a(x)$ and $a(x')$. Once we have this expression, we need to regularize the $\dfrac{1}{(x-x')^6}$ piece. This can be regularized \cite{Freedman} with the use of 
\be
\Box^2 H\left(\left(x-\xp\right)^2\right)={1\over \left(x-\xp\right)^6},
\label{regulator}
\ee
where in  $\mathbb{R}^4\diagdown \{0\}$ we can write 
\be
H\left(\left(x-\xp\right)^2\right)=-{1\over 32}{\log\, \left(\m^2\left(x-\xp\right)^2\right)\over \left(x-\xp\right)^2}.
\ee
The parameter $\mu^2$ is needed for dimensional reasons and it corresponds to the renormalization group scale \cite{Freedman}.  With \eqref{regulator} we can regularize \eqref{twopoint} the renormalized two-point function
\be
\Delta^{(2)r}(x,\xp)={\l^2\over 384 \pi^6} {1\over 32 a^3(x) a^3(\xp)}\Box^2 {\log\, \left(\m^2\left(x-\xp\right)^2\right)\over \left(x-\xp\right)^2}.
\ee
This contribution clearly breaks conformal invariance; this breaking is summarized in the corresponding broken Ward identities (the renormalization group equations) to be introduced momentarily.
\par
To see another example, let us look at the one-loop four-point amplitude where we again use the expression in \cite{Freedman} with the changes needed for CF spacetimes
\bea
&\Gamma(x_1,x_2,x_3,x_4)=-\l\,\d(x_2-x_1)\d(x_3-x_1)\d(x_4-x_1)+{\l^2\over 32\pi^4}\bigg\{\d(x_1-x_2)\d(x_3-x_4) {1\over a^2(x_1)a^2(x_3) (x_1-x_3)^4}+\nonumber\\
&+\d(x_1-x_3)\d(x_2-x_4){1\over a^2(x_1)a^2(x_2)(x_1-x_2)^4}+\d(x_2-x_3)\d(x_4-x_1){1\over a^2(x_2)a^2(x_4)(x_2-x_4)^4}\bigg\}.
\eea
This amplitude can be regulated using the identity (valid again in $\mathbb{R}^4\diagdown \{0\}$)
\be
{1\over x^4}=-{1\over 4}\Box {\log\, x^2\m^2\over x^2}.
\ee
This is enough to ensure  the existence of a Fourier transform. Actually
\be
-{1\over 4}\int d^4 x\, e^{i k x}\,\Box {\log\, x^2\m^2\over x^2}={\pi^2\over k^2}\log\,{k^2\over \m^2}.
\ee
The renormalized four-point function then simply reads
\bea
&\Gamma^{r}(x_1,x_2,x_3,x_4)=-\l\,\d(x_{21})\d(x_{31})\d(x_{41})-{\l^2\over 128\pi^4}\bigg\{\d(x_{12})\d(x_{34}) {1\over a^2(x_1)a^2(x_3)} \Box {\log\, x_{13}^2\m^2\over x_{13}^2}+\nonumber\\
&+\d(x_{13})\d(x_{24}){1\over a^2(x_1)a^2(x_4)}\Box {\log\, x_{14}^2\m^2\over x_{14}^2}+\d(x_{23})\d(x_{14}){1\over a^2(x_2)a^2(x_4)}\Box {\log\, x_{24}^2\m^2\over x_{24}^2}\bigg\},
\eea
where $x_{ij}\equiv x_i-x_j$. This four-point function obeys a renormalization group equation summarizing the broken Ward identity of scale invariance valid for any CF spacetime, namely, 
\be
\m\\{d \over d\m} \Gamma(x_1,x_2,x_3,x_4)={3\l^2\over 16\pi^2}\,{1\over a^4(x_1)}\,\d(x_{21})\d(x_{31})\d(x_{41}).
\ee
This corresponds in flat spacetime to a beta function
\be
\b(\l)={3\l\over 16\pi^2}
\ee
Nevertheless, in non-flat (but Weyl-flat) spacetimes the interpretation of the Ward identity is more involved owing to the factors of the scale factor in front of the second member.
\par
Finally, the fermion self-energy owing to a Yukawa coupling $g$, is equally easy to compute in CF spacetimes from the flat spacetime result
\be
\Sigma(x,\xp) =g \, S(x,\xp)\Delta(x,\xp)=-{g\over 8\pi^4}\g^\m {(x-\xp)_\m\over a(x)^{5/2} a(\xp)^{5/2} (x-\xp)^6}.
\ee
The tricks of the trade in \cite{Freedman} tell us that
\be
{x^\m\over x^6}=-{1\over 4}\pd^\m {1\over x^4}=\pd^\m\left({1\over 16} \Box {\log\,\m^2 x^2\over x^2}\right),
\ee
in such a way that
\be
\Sigma^r (x,\xp)= -{1\over 128 \pi^4 a(x)^{5/2} a(\xp)^{5/2}}\slashed{\pd}\Box{\log\,\m^2 (x-\xp)^2\over (x-\xp)^2}.
\ee
It is clear then  that we can explicitly compute the lowest terms in the loop expansion of the breaking of Weyl symmetry by interactions for CF spacetimes as we can use the position space techniques with a trivial generalization (introducing the conformal distance) of flat spacetime computations. 

\section{Conclusions}
In this paper, we have shown that in Weyl-flat spacetimes (that is conformally flat spacetimes) we can define conformal fields quite similar to the ones in flat spacetime. The set of Weyl-flat spacetimes is quite large, including in particular all FRW cosmological models. De Sitter and anti-de Sitter are just particular cases corresponding to constant scalar curvature. In this sense, our work is a significant generalization of the work in \cite{Allen1,Allen2,Shore}.
\par
We find that the Conformal Killing Vectors are {\em exactly} the same as the ones in flat spacetime when expressed in terms of conformally flat coordinates. Their intrinsic definition can be easily inferred from this.
\par
The guiding principle for our work has been the systematic use of a novel definition of distance. Indeed, the basic variable that allows simplifying the equations of motion is what we have dubbed as the {\em conformal distance}, a biscalar that behaves under conformal Killing transformations in the same way as the metric itself. This is a definition that generalizes properties that are shared by the usual two-point distance in flat spacetime.
\par
Scalar conformal field theories are then obtained when the Weyl invariant non-minimal coupling is used. This fact uncovers a fascinating interplay between Weyl invariance and conformal invariance. This generalizes previously known relationships between Weyl invariance and the conformal group (cf. \cite{conformalweyl}).
\par
Scalar and spinorial Green functions in all those spacetimes can then be obtained from the flat space ones simply by replacing the Euclidean distance by the conformal distance. The way that Weyl invariance is broken when interactions are included has also been explicitly worked out, with the generalization of the use of DR techniques of \cite{Freedman}.
\par 
Most nontrivial CFT in flat space are supersymmetric ones. The next step would be to try and generalize those theories to conformally flat spacetimes. Actually, it is known  \cite{Kehagias,Cassani} that it is precisely in conformally flat spacetimes where it is possible to define rigid supersymmetric theories. Work is currently going on in this direction, and we hope to report on it in due time.
\par
We end with a related point. Even theories that are conformal in flat space (such as N=4 super Yang Mills) get gravitational contributions to the beta function when considered in a non-trivial gravitational background. It would be quite interesting to examine what restrictions conformal invariance implies for those gravitational contributions.

\section{Acknowledgments}

This work has received funding from the Spanish Research Agency (Agencia Estatal de Investigacion) through the grant IFT Centro de Excelencia Severo Ochoa SEV-2016-0597, and the European Union's Horizon 2020 research and innovation programme under the Marie Sklodowska-Curie grants agreement No 674896 and No 690575. We have also been partially supported by FPA2016-78645-P(Spain). RSG is supported by the Spanish FPU Grant No FPU16/01595.
\appendix
\section{Constant curvature spacetimes}\label{A}
A particular instance of conformally flat spacetimes is given by spacetimes of constant curvature. 
Some of the formulas in the main text can be made even more explicit in this case. In particular, the correspondence between CKV in flat spacetime and in the constant curvature spacetime can be made explicit (in the general case it depends on the detailed form of the functions $a(x)$). We present some detailed results in this appendix.
\par
In our conventions constant curvature spacetimes fulfill
\be
R_{\a\b\g\d}={R\over n(n-1)}\left(g_{\a\g}g_{\b\d}-g_{\a\d}g_{\b\d}\right),
\ee
where the scalar curvature is given in terms of the cosmological constant by
\be
R\equiv-{2 n\over n-2}\l.
\ee
A theorem by Lichnerowicz \cite{Lichnerowicz} ensures that precisely in these spacetimes there exist Killing spinors (eigenvectors of Dirac's operator).
In order to be specific we shall particularize in most of the formulas to de Sitter space $dS_n$
\be
ds^2=L^2\,{dz^2-\sum_{i=1}^{i=n-1} dy_i^2\over z^2},
\ee
which in our conventions has negative curvature but positive cosmological constant
\be
R=-{n(n-1)\over L^2}\equiv -{2 n \over n-2}\,\l.
\ee
Let us study in some detail the group generated by all CKV. Although it is mathematically isomorphic to the flat space conformal group, the correspondence is not one to one between the generators. Let us recall the definition of CKV
\be
\nabla_\m \xi_\n+\nabla_\n\xi_\m={2\over n} \nabla_\l \xi^\l\,g_{\m\n}\equiv w(x)g_{\m\n},
\ee
In de Sitter space, in particular, we have the formula
\be
w_{dS}=w_{flat}-{2\over z}\xi^z
\ee
To be specific
\bi
\item  Dilatations are implemented by a real Killing vectors, because
\be
w=2\l- {2\over z}\l z=0.
\ee
\item On the other hand, translations obey
\be
\nabla_\m \xi^\m=-{n\over z} a^z,
\ee
so that the $n-1$ translations with $a^z=0$ are real Killings, and $a=(1,\vec{0})$ is a CKV with
\be
w(x)=-{2\over z}.
\ee
\item Lorentz transformations obey
\be
\nabla_\m\xi^\m=-{n\over z}\omega^z\,_j y^j,
\ee
so that the ${(n-1)(n-2)\over 2}$ transformations with $\omega^z\,_j=0$ are real Killings
and the $n-1$ remaining ones  are CKV with
\be
w(x)=-{2\over z}\omega^z\,_j y^j
\ee
\item Finally the n special conformal transformations (SCT) yield
\be
w(x)=2\,x^2\,{\b^z\over z}
\ee
These transformations are then implemented by CKV except when $\b^z=0$; and there are $n-1$ of those.
\ei
Altogether there are ${n(n+1)\over 2}$ Killings and $(1+n)$ which are only CKV. Let us summarize the classification of all  generators into those which are real Killinsg and those that are conformal Killings in the following table.
\vspace{1cm}

\begin{center}
	
	\begin{tabular}{|c|| c || c| }
		\hline 
		Killing&Flat & de Sitter  \\ \hline                        
		Translations &$w=0$&$w=-{2 a^z\over z}$ \\
		Dilatations&$2\l$ & $w=0$\\
		Lorentz&$w=0$&$w=-{2 \omega^z\,_j y^j\over z}$\\
		Special CT&$w=4 \b.x$&$w={2 x^2\over z} \b^z$\\
		\hline  
	\end{tabular}
	
\end{center}


\vspace{1cm}

\par

%
%

Let us note that the geodesic distance and the conformal distance are not equivalent, as can be seen in this particular case. It is well known that the geodesic distance in de Sitter space between timelike separated points obeys
\be
\cosh\,{s\over L}={z^2+(z^\prime)^2-\left({y_i^\prime}-{y_i}\right)^2\over 2 z z^\prime}=1+{\Delta z^2-\Delta y^2\over 2 z z^\prime}
\label{geodesicdistance}
\ee
as in \cite{Schomblond}. This distance is not equivalent to the conformal distance definition \eqref{C} that in this case yields
\be
C(x,\xp)_{dS} = \frac{L^2}{z z'} (\Delta z^2-\Delta y^2).
\label{conformaldS}
\ee
As it is mentioned in the text, we can see that the world distance can be expressed as a function of the conformal distance. 


\newpage
\begin{thebibliography}{99}
	\bibitem{Rychkov}
	S.~Rychkov,
	``EPFL Lectures on Conformal Field Theory in D$\geq$ 3 Dimensions,''
	doi:10.1007/978-3-319-43626-5
	arXiv:1601.05000 [hep-th].
	\bibitem{Osborn}
	H. Osborn,
	``Lectures on conformal field theories,"
	(preprint 2018)
	\bibitem{Petrov}
	A.~Z.~Petrov,
	``The Classification of spaces defining gravitational fields,''
	Gen.\ Rel.\ Grav.\  {\bf 32} (2000) 1665.
	doi:10.1023/A:1001910908054
	\bibitem{Eisenhart}
	L.P. Eisenhart,
	``Riemannian Geometry"
	(Princeton)
	\bibitem{Backovsky}
	P. Backovsky and J. Niederle,
	``On classification of Conformally flat spaces"
	Check J. Phys, 47(1996),N0 10,p1001
	\bibitem{Synge} 
	J.~L.~Synge,
	``Relativity: The General theory,''
	North-Holland, Amsterdam, 1960
	\bibitem{DeWitt} 
	B.~S.~DeWitt,
	``Dynamical theory of groups and fields,''
	Conf.\ Proc.\ C {\bf 630701}, 585 (1964)
	[Les Houches Lect.\ Notes {\bf 13}, 585 (1964)].
	\bibitem{Lichnerowicz}
	A.~Lichnerowicz,
	``Spin manifolds, Killing spinors and universality of the Hijazi inequality,''
	Lett.\ Math.\ Phys.\  {\bf 13} (1987) 331.
	doi:10.1007/BF00401162
	\bibitem{Lichne}
	A.~Lichnerowicz,
	``Propagators, commutators and anti-commutators in general relativity,''
	Gen.\ Rel.\ Grav.\  {\bf 50} (2018) no.11,  145.
	doi:10.1007/s10714-018-2433-x\\
	\bibitem{Shore}
	H.~Osborn and G.~M.~Shore,
	``Correlation functions of the energy momentum tensor on spaces of constant curvature,''
	Nucl.\ Phys.\ B {\bf 571} (2000) 287
	doi:10.1016/S0550-3213(99)00775-0
	[hep-th/9909043].
	\bibitem{Jack}
	I.~Jack and H.~Osborn,
	``Background Field Calculations in Curved Space-time. 1. General Formalism and Application to Scalar Fields,''
	Nucl.\ Phys.\ B {\bf 234} (1984) 331.
	doi:10.1016/0550-3213(84)90067-1
	\bibitem{Freedman}
	D.~Z.~Freedman, K.~Johnson and J.~I.~Latorre,
	``Differential regularization and renormalization: A New method of calculation in quantum field theory,''
	Nucl.\ Phys.\ B {\bf 371} (1992) 353.
	doi:10.1016/0550-3213(92)90240-C
	\bibitem{Gracia}
	J.~M.~Gracia-Bond\'ia, H.~Guti\'errez-Garro and J.~C.~V\'arilly,
	``Improved EpsteinÐGlaser renormalization in x-space versus differential renormalization,''
	Nucl.\ Phys.\ B {\bf 886} (2014) 824
	doi:10.1016/j.nuclphysb.2014.07.018
	[arXiv:1403.1785 [hep-th]].
	\bibitem{Allen1}
	B.~Allen and T.~Jacobson,
	``Vector Two Point Functions in Maximally Symmetric Spaces,''
	Commun.\ Math.\ Phys.\  {\bf 103} (1986) 669.
	doi:10.1007/BF01211169
	\bibitem{Allen2}
	B.~Allen and C.~A.~Lutken,
	``Spinor Two Point Functions in Maximally Symmetric Spaces,''
	Commun.\ Math.\ Phys.\  {\bf 106} (1986) 201.
	doi:10.1007/BF01454972
	\bibitem{conformalweyl}
	E.~Alvarez, J.~Anero and R.~Santos-Garcia,
	``Conformal invariance versus Weyl invariance,''
	arXiv:1903.05653 [hep-th].\\
	\bibitem{Kehagias}
	A.~Kehagias and J.~G.~Russo,
	``Global Supersymmetry on Curved Spaces in Various Dimensions,''
	Nucl.\ Phys.\ B {\bf 873} (2013) 116
	doi:10.1016/j.nuclphysb.2013.04.010
	[arXiv:1211.1367 [hep-th]].
	\bibitem{Cassani}
	D.~Cassani and D.~Martelli,
	``Supersymmetry on curved spaces and superconformal anomalies,''
	JHEP {\bf 1310} (2013) 025
	doi:10.1007/JHEP10(2013)025
	[arXiv:1307.6567 [hep-th]].
	\bibitem{Schomblond}
	C. Scomblond and P. Spindel,
	``Unicity Conditions of the Scalar Field Propagator Delta(1) (x,y) in de Sitter Universe,''
	Ann Ins. Henri Poincare, XXV n1 (1976) 67
	
	
	
	
	%
	%
	
	%
	
	
	
	
	
	
	
	
	
	
	
	
	
	
	
	
	
	
\end{thebibliography}
\end{document}